\documentclass[journal]{IEEEtran}

\usepackage{lineno} % 这个包就是latex自带的添加行号的包了
\usepackage{amsmath,amsfonts}
\usepackage{array}
\usepackage{subcaption}
\usepackage{textcomp}
\usepackage{stfloats}
\usepackage{url}
\usepackage{verbatim}
\usepackage{graphicx}
\usepackage{cite}
\usepackage[bookmarksnumbered=true]{hyperref}
\hypersetup{
hidelinks,
colorlinks=true,
linkcolor=blue,
citecolor=blue,
urlcolor=black
}
\usepackage[linesnumbered,ruled,vlined]{algorithm2e}
\usepackage{bbm} % 字体bbl,空心字母，指示复数空间

\usepackage{amsthm} % theorem,lemma,proof环境

\usepackage{hyperref}
\usepackage{xcolor}
\hypersetup{
    colorlinks=true,      % 启用颜色超链接
    urlcolor=blue        % 设置URL超链接颜色为蓝色
}

\usepackage{enumerate} % 有序编号

\begin{document}

\newcounter{probcounter}
\newcommand{\refporbcounter}[1]{\refstepcounter{probcounter}\theprobcounter \label{#1}} % 新建命令，在正文中插入可引用的计数器，但是不能直接在公式环境中用

\title{Wireless Hallucination in Generative AI-enabled Communications: Concepts, Issues, and Solutions}

\author{
Xudong Wang, 
Jiacheng Wang,
Lei Feng,
Dusit Niyato,~\IEEEmembership{Fellow,~IEEE,}\\
Ruichen Zhang,
Jiawen Kang,
Zehui Xiong,
Hongyang Du,
Shiwen Mao,~\IEEEmembership{Fellow,~IEEE}

\thanks{\textit{Corresponding author: Lei Feng.}}
\thanks{
X. Wang and L. Feng are with the State Key Laboratory of Networking and Switching Technology, Beijing University of Posts and Telecommunications, Beijing, China, 100876 
(e-mail: xdwang@bupt.edu.cn, fenglei@bupt.edu.cn). J. Wang, D. Niyato, and R. Zhang are with the College of Computing and Data Science,
Nanyang Technological University, Singapore (e-mail: jiacheng.wang@ntu.edu.sg, dniyato@ntu.edu.sg, ruichen.zhang@ntu.edu.sg). J.~Kang is with the School of Automation, Guangdong University of Technology, Guangzhou 510006, China (e-mail: kavinkang@gdut.edu.cn). Z. Xiong is with the Pillar of Information Systems Technology and Design, Singapore University of Technology and Design, Singapore (e-mail: zehui\_xiong@sutd.edu.sg). H. Du is with the Department of Electrical and Electronic Engineering, University of Hong Kong, Pok Fu Lam, Hong Kong SAR, China (e-mail: duhy@eee.hku.hk). S. Mao is with the Department of Electrical and Computer Engineering, Auburn University, Auburn, USA (e-mail: smao@ieee.org).}
% \thanks{}
% \thanks{}
% \thanks{}
% \thanks{}
}



\maketitle

\begin{abstract}

% Generative AI (GenAI) is driving the intelligence of wireless communications. However, due to data deficiencies, stochastic generation, and dynamic environments, GenAI may produce channel information or optimization strategies that violate physical laws, leading to wireless hallucinations. This issue results in invalid channel information, spectrum waste, and reduced communication reliability, yet it remains underexplored in a systematic manner. To address this gap, this paper provides a comprehensive review of wireless hallucinations in GenAI-driven communications, with a particular emphasis on hallucination mitigation. Specifically, we introduce the concept of wireless hallucinations for the first time, analyze their underlying causes based on the GenAI workflow, and propose mitigation strategies at the data level, model level, and post-generation level. Then, we systematically analyze representative hallucination scenarios in GenAI-enabled communications along with their corresponding mitigation strategies. Finally, we propose a novel hybrid mitigation strategy for GenAI-based channel estimation to reduce potential hallucination effects. At the data level, we release a channel estimation hallucination dataset and employ GAN-based data augmentation. Additionally, we incorporate attention mechanisms and large language models (LLMs) to enhance both training and inference performance. Experimental results demonstrate that the proposed hybrid mitigation strategy reduces the normalized mean square error (NMSE) by 0.19, effectively reducing wireless hallucinations.
Generative AI (GenAI) is driving the intelligence of wireless communications.
Due to data limitations, random generation, and dynamic environments, GenAI may generate channel information or optimization strategies that violate physical laws or deviate from actual real-world requirements. 
We refer to this phenomenon as wireless hallucination, which results in invalid channel information, spectrum wastage, and low communication reliability but remains underexplored.
To address this gap, this article provides a comprehensive concept of wireless hallucinations in GenAI-driven communications, focusing on hallucination mitigation. Specifically, we first introduce the fundamental, analyze its causes based on the GenAI workflow, and propose mitigation solutions at the data, model, and post-generation levels. Then, we systematically examines representative hallucination scenarios in GenAI-enabled communications and their corresponding solutions. Finally, we propose a novel integrated mitigation solution for GenAI-based channel estimation. At the data level, we establish a channel estimation hallucination dataset and employ generative adversarial networks (GANs)-based data augmentation. Additionally, we incorporate attention mechanisms and large language models (LLMs) to enhance both training and inference performance. Experimental results demonstrate that the proposed hybrid solutions reduce the normalized mean square error (NMSE) by 0.19, effectively reducing wireless hallucinations.
% Generative AI (GenAI) enhances wireless communication intelligence but may generate channel information or optimization strategies that violate physical laws or deviate from requirements due to data limitations, stochastic generation, and dynamic environments—a phenomenon we term wireless hallucination. This issue results in invalid channel information, spectrum waste, and reduced communication reliability but remains underexplored. 
\end{abstract}

% \begin{IEEEkeywords}
% Generative Artificial Intelligence, Wireless Hallucination, Diffusion Model, Channel Estimation
% \end{IEEEkeywords}

\section{Introduction}

% \textbf{First paragraph:} Introduce generative AI technology, its characteristics, and advantages.
% Generative AI (GenAI) is an artificial intelligence technology that generates new content, such as text, images, or audio, by learning from existing data. Examples include OpenAI's GPT series, which has garnered widespread attention and research in recent years. GenAI captures patterns and structures from large training datasets, enabling it to generate content with similar styles or structures, thereby enhancing creativity and decision-making in practical applications. With its demonstrated advantages in customization and diversity, GenAI has become a vital tool for driving innovation across various fields, including drug discovery, game development, and more.

% \textbf{Second paragraph:} Explain the growing attention to GenAI technology in the field of wireless communications. Review related work that applies GenAI techniques to address issues in wireless communication systems.
Generative AI (GenAI) technology has garnered widespread attention and research interest in recent years. By learning from vast amounts of training data, GenAI captures patterns and structures to generate content with similar styles or information. Notably, thanks to its powerful ability to model complex scenarios and dynamically adapt, GenAI is increasingly being applied in wireless communications, such as channel estimation, interference management, and resource allocation. By leveraging its generative capabilities, GenAI-driven models can facilitate efficient network optimization, enhance adaptability to dynamic wireless environments, and improve spectrum utilization through predictive and generative reasoning~\cite{khoramnejad2025generative}. The integration of GenAI into wireless networks enables autonomous modeling of complex wireless environments, paving the way for intelligent wireless networks capable of self-learning and adaptive decision-making.
Despite significant progress, the application of GenAI in wireless communications still faces challenges, one of which is known as ``wireless hallucination." Hallucination is a widely discussed issue in GenAI applications especially large language models (LLMs), referring to the generation of false, inaccurate, or entirely fabricated content by GenAI models~\cite{chakraborty2025hallucination}. With subtle differences from errors, which typically stem from computational mistakes or incorrect input data, hallucinations involve the model producing seemingly plausible but unverifiable or entirely fictional information. While errors can often be corrected through debugging or data refinement, hallucinations require more advanced mitigation techniques, such as fact-checking or retrieval-augmented generation. The causes of hallucination are primarily associated with the model's training methodology, limited training data, and the probabilistic nature of generative models~\cite{huang2024survey}. This challenge also exists in GenAI-enabled wireless communications and have adverse impacts, such as resource wastage, degraded user quality of experience (QoE), and even security risks to wireless networks. For example, in RF signal generation, existing GenAI models struggle to accurately capture the time-frequency characteristics and hardware-specific traits of the devices, leading to RF signals that appear structurally reasonable but deviate from real-world physical constraints. Unlike conventional errors, these wireless hallucinations introduce fictitious spectral components or unrealistic distortions that do not correspond to any valid transmission, thereby weakening the robustness of classifiers and reducing wireless sensing capabilities~\cite{chi2024rf}. Actually, existing work on GenAI-empowered wireless communications has not systematically addressed this critical challenge, nor proposed comprehensive solutions to tackle it.
To this end, this paper is the first to investigate hallucinations in GenAI-enabled wireless communications and propose mitigation strategies. We begin by reviewing hallucinations in GenAI models, defining wireless hallucination, categorizing its types, and identifying its causes. Then, we propose mitigation strategies at the data, model, and post-generation levels. Additionally, we analyze existing GenAI-enabled wireless communication schemes, focusing on their potential hallucination causes and employed mitigation methods. Since most studies adopt a single mitigation approach with limited performance, we introduce a GenAI-enabled channel estimation scheme as a case study and propose an integrated mitigation solution to enhance estimation accuracy. The main contributions are summarized as follows:

\begin{itemize}
% 缩短前
    % \item We provide a comprehensive overview of generalized hallucination in GenAI models and introduce the definition of wireless hallucination from the perspective communications. Wireless hallucination is categorized into three primary types. By systematically analyzing the workflow of GenAI-enabled communications, we reveal the causes of wireless hallucination in detail. More importantly, we propose mitigation strategies from data-level, model-level, and post-generation level.
    \item We provide a comprehensive overview of generalized hallucination in GenAI models and define wireless hallucination from the perspective of communications. By systematically analyzing the workflow of GenAI-enabled communications, we reveal its causes and propose solutions at the data, model, and post-generation levels.
    % 缩短前
    % \item We review existing GenAI-enabled wireless communication paradigms from the perspectives of the physical layer, data link layer, and network layer, such as radio map construction and GenAI-enabled semantic communication. Special emphasis is placed on analyzing the causes of wireless hallucination and exploring corresponding hallucination mitigation strategies.
    \item We review existing GenAI-enabled wireless communication paradigms from the physical layer, data link layer, and network layer, such as radio map construction and GenAI-enabled semantic communication. Special focus is given to identifying the causes of wireless hallucination and exploring mitigation strategies.
    % 缩短前
    % \item We propose a novel hybrid mitigation approach to reduce wireless hallucination in GenAI-enabled uplink channel estimation. Generative adversarial networks (GANs) address the imbalance in channel data distribution, while an attention mechanism enhances the reverse denoising process of the diffusion model. Additionally, an LLM-enhanced mixture of experts (MoE) analyzes user information to select optimal experts for channel estimation. Experimental results show that our approach reduces normalized mean squared error (NMSE) by $0.19$, effectively reducing wireless hallucinations.
    \item We design an efficient integrated strategy to reduce wireless hallucination in GenAI-enabled uplink channel estimation. More specifically, generative adversarial networks (GANs) are used to balance channel data distribution, while an attention mechanism enhances the reverse denoising of the diffusion model. Additionally, an LLM-enhanced mixture of experts (MoE) is developed to select optimal experts based on user information. Experimental results show that the proposed integrated approach reduces normalized mean squared error (NMSE) by $0.19$, effectively reducing wireless hallucinations.
\end{itemize}
% Paragraph 4: (Contribution) In this paper, we propose...
% \begin{itemize}
%     \item We.
%     \item We.
%     \item T.
% \end{itemize}

% \section{Application of Matching Generation}
\section{Hallucination In Wireless Communications}

\subsection{Overview of Basic Hallucination in GenAI Models}

The hallucinations and their underlying causes in common GenAI models are discussed based on their principles, including Variational Autoencoders (VAEs), GANs, generative diffusion models (GDMs), and Transformers:

\subsubsection{Variational Autoencoders} VAEs~\cite{10874185} compress data into latent space representations using an encoder, then decode samples from the latent space to generate new data resembling the original input. However, over-simplification of the latent space and random sampling can introduce ambiguity and detail loss, leading to hallucinations. A typical example is contradictory sentences in descriptive text generated by VAEs.

\subsubsection{Generative Adversarial Networks} GANs~\cite{9925625} consist of a generator and a discriminator that compete to realistically mimic data distributions. Mode collapse and generator reliance on shortcuts can result in unrealistic or repetitive outputs. For instance, hallucinations may manifest as faces with multiple pupils or missing ears in generated images.

\subsubsection{Generative Diffusion Models} GDMs~\cite{wang2024generative} gradually add noise to data and learn to reverse the process, generating high-quality new samples. However, error accumulation and an imbalance between global structure and fine details during iterative denoising can produce artifacts. Such hallucinations may appear as sudden pitch changes in generated music.

\subsubsection{Transformers} Transformers~\cite{chi2024rf} use self-attention mechanisms to model global relationships in input data, generating new sequences by predicting the next element. However, over-reliance on learned correlations and attention biases can result in outputs that are plausible but factually incorrect or inconsistent. This can lead to hallucinations such as motion trajectories in generated videos that defy physical laws.

\begin{figure*}
    \centering
    \includegraphics[width=0.8\linewidth]{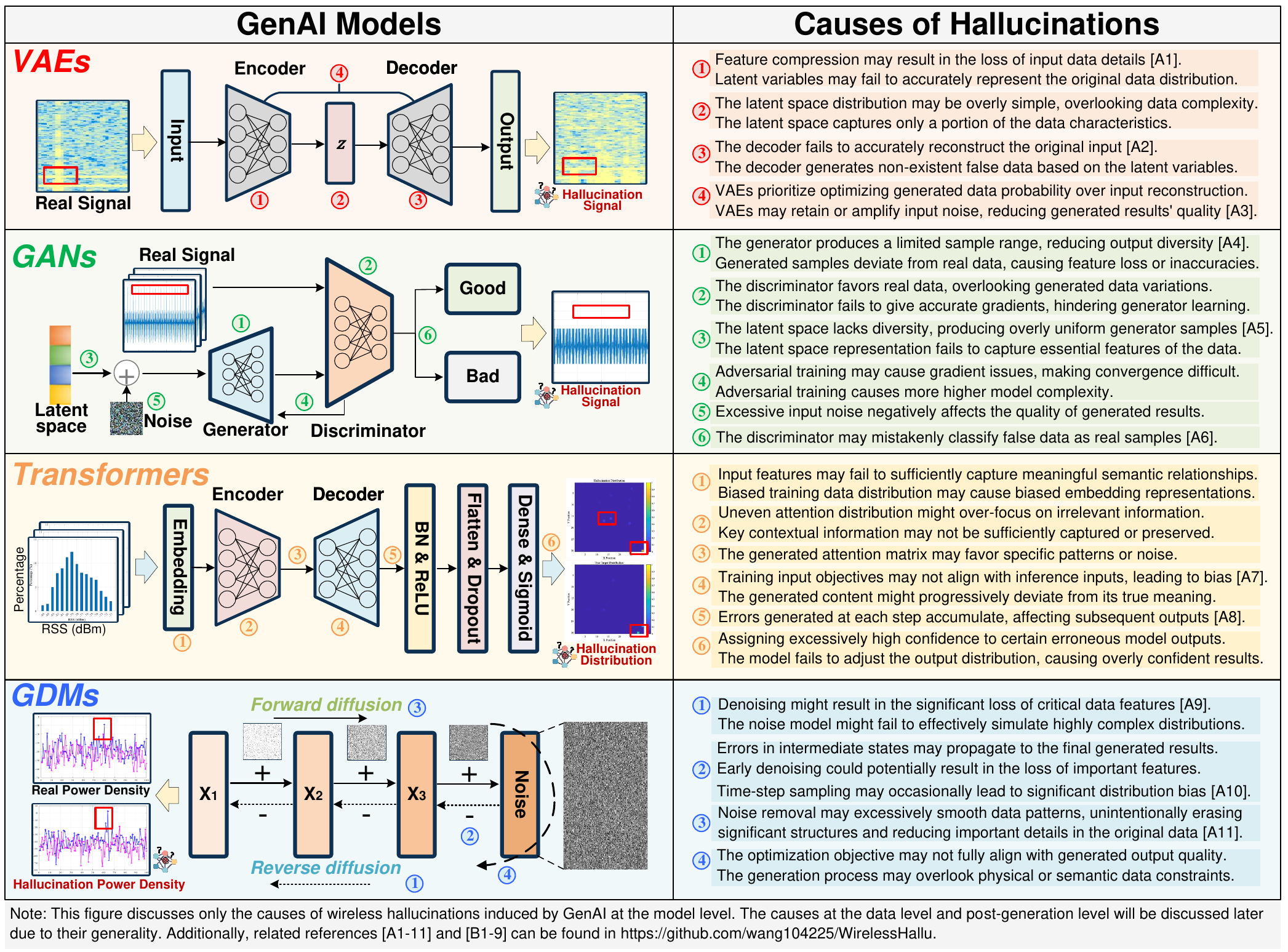}
    \caption{\small{A summary of different types of GenAI models, potential causes and manifestations of hallucination.}}
    \label{Figure1_GenAI_model_Hallucination}
\end{figure*}

We summarize the principles of different GenAI models, and illustrate the potential causes and manifestations of hallucinations within each model, shown as Fig.~\ref{Figure1_GenAI_model_Hallucination}.

\subsection{Definition of Wireless Hallucination}

\textit{Wireless hallucination refers to the phenomenon where GenAI models in wireless communication tasks, produce results that seem reasonable but are in fact inaccurate, inapplicable, or even incorrect under actual physical environments or technical constraints.} This phenomenon parallels hallucinations in natural language models, where outputs may appear valid yet fail factual verification~\cite{ji2023survey}. In wireless communications, such hallucinations manifest as fabricated, distorted, or logically flawed representations of channels, signals, protocols, or resource allocation schemes, potentially degrading system performance or rendering solutions infeasible. Wireless hallucinations primarily affect the physical and network layers and can be categorized into the following types:

\begin{itemize}
    % 缩短前
    % \item \textbf{Constraint Violation Wireless Hallucination:} This describes a situation where the content generated by the GenAI model does not conform to the theoretical or technical constraints of wireless communication. For example, with a total spectrum of $20~\mathrm{MHz}$, the GenAI model allocates $25~\mathrm{MHz}$ to a user, exceeding the available bandwidth.
    \item \textbf{Constraint Violation Wireless Hallucination:} This occurs when a GenAI model generates content that violates theoretical or technical constraints in wireless communication. For instance, if the total spectrum is $20~\mathrm{MHz}$, but the model allocates $25~\mathrm{MHz}$ to a user, it exceeds the available bandwidth.
    % 缩短前
    % \item \textbf{Fabricated Output Wireless Hallucination:} This describes a situation where the GenAI model generates seemingly realistic but entirely unsuitable optimization solutions or content. For example, the GenAI model creates a new channel fading model claiming better simulation of high-density urban environments, but this model doesn't exist in academia or industry.
    \item \textbf{Fabricated Output Wireless Hallucination:} This occurs when a GenAI model generates seemingly realistic but entirely unsuitable optimization solutions or content. For example, it may create a new channel fading model for high-density urban environments that does not exist in academic literature or industry standard.
    % 缩短前
    % \item \textbf{Context Detachment Wireless Hallucination:} This describes a situation where the content generated by GenAI is irrelevant to the task objectives or the context of the problem, deviating from the actual requirements. For example, when generating channel simulation data for urban environments, the GenAI model produces results based on a rural scenario, with excessively long propagation distances and low multipath density.
    \item \textbf{Context Detachment Wireless Hallucination:} This occurs when GenAI generates content deviating from task objectives, problem context or the actual requirements. For instance, when simulating urban channels, GenAI may incorrectly use rural parameters, resulting in overly long propagation distances and low multipath density.
\end{itemize}

\subsection{Causes of Wireless Hallucination}

% 缩短前
% Before analyzing the causes of wireless hallucinations, we first review the workflow of GenAI models in wireless communications. First, the system needs to collect and preprocess a large amount of real-time data, such as Channel State Information (CSI), network traffic, and user locations. Then, the data undergoes preprocessing and feature extraction to ensure it meets the input requirements of the GenAI model. Next, based on different problem requirements, the system inputs the data into the appropriate GenAI model, for training, guiding the model to converge towards predefined optimization goals, such as system throughput, latency, and energy efficiency. Finally, the trained GenAI model generates corresponding decisions or content based on the real-time input signals or network status data. Based on the above analysis, we summarize the potential causes leading to wireless hallucinations.
Before analyzing the causes of wireless hallucinations, we review the GenAI workflow in wireless communications. First, the system collects and preprocesses real-time data, including Channel State Information (CSI), network traffic, and user locations. This data then undergoes preprocessing and feature extraction to meet the GenAI model’s input requirements. Next, based on the problem’s needs, the system feeds the data into a suitable GenAI model for training, guiding the model to converge towards predefined optimization goals, such as throughput or latency. Finally, the trained model generates decisions or content based on real-time input signals or network status. From this workflow, we identify key factors contributing to wireless hallucinations.

\subsubsection{Insufficient or Biased Data} If the training dataset does not cover a sufficiently broad range of scenarios (e.g., different environments, interference types, channel conditions, and network), the GenAI model may only learn local or limited features, leading to errors in the generated results when applied to unseen wireless scenarios~\cite{chi2024rf,10757328}. Additionally, biased data may cause the GenAI model to favor certain outcomes during the inference process~\cite{10433140}.

\subsubsection{Inherent Randomness in GenAI Models} GenAI models, especially GDMs, begin with random noise and gradually denoise it through a reverse diffusion process to restore the target data. Variations in the initial noise can cause differences in the generated images or signals, and each denoising step may introduce randomness. As a result, even with the same input and constraints, GenAI models can produce varied outcomes, leading to some results meeting the requirements while others are unreasonable or impractical~\cite{zhang2024emergence}.

\subsubsection{Objective Setting and Task Understanding Issues} If the task or goal is unclear, the model may generate irrelevant or conflicting outputs. For instance, increasing throughput might require higher transmission power, but this could worsen interference and reduce latency efficiency. Insufficient task understanding is reflected in the GenAI model's misunderstanding or incomplete understanding of the physical, technical, and constraint conditions of wireless communication tasks. As a result, GenAI models may overlook critical constraints or context in wireless communication, leading to the generation of invalid or unsuitable solutions~\cite{10839236}.

% \subsubsection{Lack of domain knowledge}AI models still lack a deep understanding of fundamental wireless communication theory

\subsubsection{Absence of Validation Procedure} If the validation procedure is missing, the results from GenAI may not be compared or validated against actual requirements and constraints. The generated solutions may overlook these constraints, leading to outputs that violate real-world communication rules or fail to adapt to the actual environment, thus producing unrealistic results and causing wireless hallucinations~\cite{du2023user,9925625}.

\section{Mitigation Solutions for Wireless Hallucination}

% 缩短前
% The strategies to mitigate wireless hallucinations are summarized into three main aspects. In data-level solutions, data augmentation and data cleaning help enhance the expected output of the GenAI model. In model-level solutions, optimizing the model structure and improving model robustness can effectively prevent the model from generating wireless hallucinations under unstable or extreme conditions. Additionally, some post-generation validation and correction methods can effectively mitigate the propagation of hallucinations.
% 缩短后
% We summarize the wireless hallucination solutions in three main categories. Data-level solutions involve data augmentation and cleaning to enhance the model’s expected output. Model-level solutions focus on optimizing model structures and improving robustness to prevent hallucinations in unstable or extreme conditions. Post-generation validation and correction methods help reduce the propagation of hallucinations.

\subsection{Data-Level Solutions}

%缩短前
% \subsubsection{Data Augmentation}Data augmentation techniques can be used to expand the training dataset, thereby combating the complexities and variability of wireless environments and reducing hallucination outputs. For example, injecting noise, time shifts, and frequency offsets into the original signals can enhance the signals. Additionally, wireless communication simulation platforms such as NS-3 can be employed to generate rich simulation data, simulating the impact of different mobility patterns, user distributions, device types, weather conditions, and other factors on network performance. On the other hand, semi-automated annotation tools (such as data labeling tools and annotation platforms) can be used for precise labeling of network data.
\subsubsection{Data Augmentation}Data augmentation techniques help expand training datasets, addressing wireless environment variability and reducing hallucinations. For instance, injecting noise, time shifts, and frequency offsets enhances signal diversity. Simulation platforms such as NS-3 can generate rich datasets, modeling the effects of mobility patterns, user distributions, device types, and weather conditions on network performance. Additionally, semi-automated annotation tools, such as labeling platforms, can improve the accuracy of network data annotations.

% 缩短前
% \subsubsection{Data Cleaning}Anomaly detection methods can effectively identify data points that significantly deviate from the normal data distribution, which may be caused by sensor faults, network issues, or other external interferences. Techniques such as Z-score and Interquartile Range (IQR) can be used to detect outliers in the data. Additionally, clustering methods such as K-means or Density-Based Spatial Clustering of Applications with Noise (DBSCAN) can be employed to analyze the data and identify abnormal data points that do not conform to regular patterns. This is particularly useful for multi-dimensional data in wireless communications (e.g., signal strength, latency, throughput). Furthermore, by generating hash values for the data items, we can quickly check for duplicates, thereby eliminating potential computational burdens on the model caused by redundant data, which might result from transmission interruptions or repeated transmissions in wireless communication.
\subsubsection{Data Cleaning}Anomaly detection can effectively identify data deviations caused by sensor faults, network issues, or external interference. Techniques like Z-score and interquartile range (IQR) detect outliers, while clustering methods such as K-means analyzes multi-dimensional wireless data (e.g., signal strength, latency, throughput) for anomalies. Additionally, hash-based duplicate detection minimizes computational overhead from redundant transmissions.

% \subsubsection{Multi-scenario data collection}Collecting real communication data in dynamic environments

\subsection{Model-Level Solutions}

\subsubsection{Dynamic Feedback and Reinforcement Learning (RL)}RL, through the interaction between an agent and its environment, enables adjustments in each decision based on the actual network state, thereby reducing the generation of optimization strategies that do not align with the real environment, i.e., wireless hallucinations. Additionally, RL facilitates online learning to continuously adjust decisions in real-time, enabling it to adapt to changes in the wireless network and avoid generating static or unreasonable resource allocation schemes.

%考虑要不要换成attention
\subsubsection{Attention Mechanism}An attention mechanism can dynamically focus on different parts of the input data, enabling more effective information extraction and modeling, particularly in the presence of high noise or incomplete data, thereby reducing the occurrence of hallucinations. For example, in frequency-selective fading channels, attention can help the model learn frequency-domain correlations, leading to more stable estimation results and minimizing the likelihood of hallucinated channel responses.

\subsubsection{Adversarial Training}Adversarial training involves generating adversarial examples to train the model, enabling it to better handle complex, anomalous, or disruptive environments, thereby improving the model's robustness and reducing hallucinated outputs. Generative Adversarial Networks have the ability to generate perturbation signals or data, and using these adversarial examples during training can effectively help the GenAI model enhance its adaptability to irregular data.
% \subsubsection{Regularization Techniques}Regularization is a technique used to prevent model overfitting and mitigate wireless hallucinations in GenAI models. L2 regularization penalizes the size of the weights, forcing the model to be more concise and robust, thus avoiding unnecessary overfitting to the training data, which effectively prevents the optimization solutions generated by the model from deviating from the actual environment.

\subsection{Post-Generation Solutions}

\subsubsection{Interactive AI (IAI)} IAI verification integrates wireless expertise with machine automation, enabling real-time feedback and adjustments to generated results. LLMs, with their strong natural language understanding and generation capabilities, serve as effective tools for interpreting and validating communication protocols and constraints. LLMs can identify issues in generated network management policies and configuration, e.g., signal interference, congestion, and even network attacks. As such, real-time feedback can be achieved to detect, prevent, and correct hallucination.
\subsubsection{Mixture of Experts} The MoE architecture enhances accuracy and adaptability by integrating multiple expert models, selecting the most suitable one for specific tasks while maintaining efficiency. Through a gating mechanism, MoE ensures that GenAI-generated solutions are validated across the appropriate expert domains. Additionally, when GenAI proposes multiple implementation approaches, the MoE framework evaluates and optimizes them by leveraging the appropriate expert, ultimately identifying the best solution and mitigating wireless hallucinations.

\section{Wireless Hallucination and Mitigation in GenAI-Enabled Communications}

\begin{figure*}
    \centering
    \includegraphics[width=0.8\linewidth]{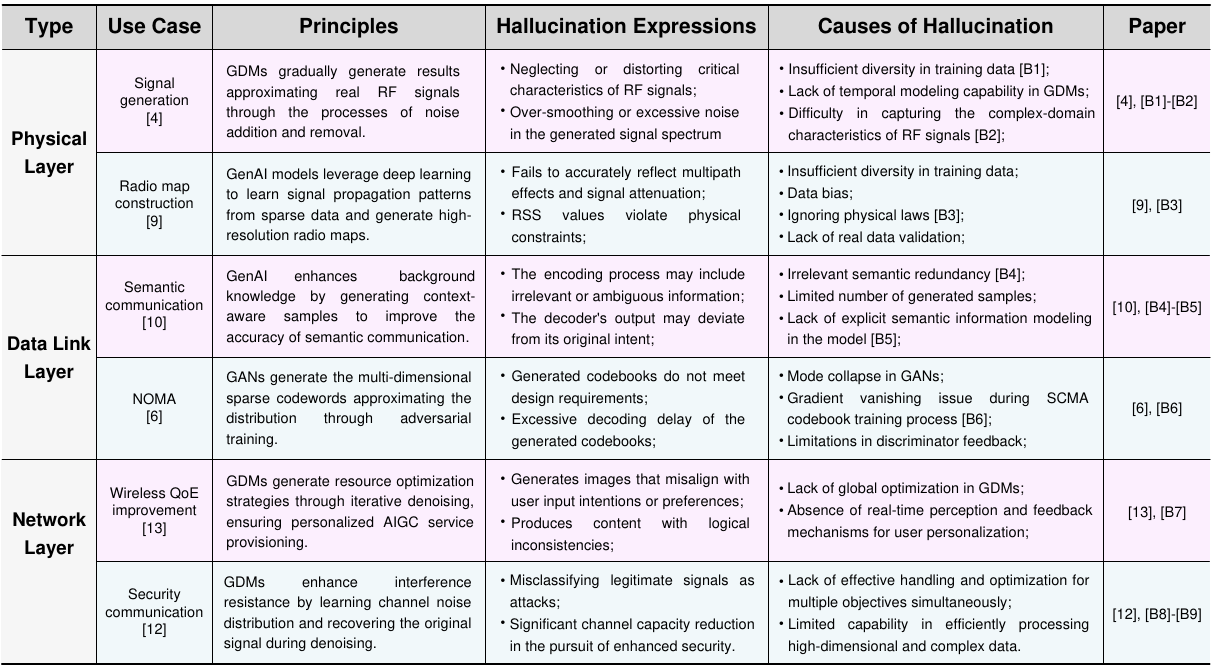}
    \caption{\small{The principles, hallucination manifestations, and causes of hallucination in the application of GenAI to wireless communications from the perspectives of the physical layer, data link layer, and network layer.}}
    \label{Table1_Hallucination_different_GenAI}
\end{figure*}

We summarize the principles, hallucination expressions, and causes of hallucination in GenAI-enabled wireless communications from the physical layer, data link layer, and network layer, as illustrated in Fig.~\ref{Table1_Hallucination_different_GenAI}.

\subsection{Physical Layer Perspective}

\subsubsection{GenAI-enabled Signal Generation}

% Some studies suggest that GenAI models can play a role in wireless data augmentation, signal detection, and time-series prediction. However, due to the dynamic temporal characteristics, frequency domain features, and complex-valued nature of radio frequency (RF) signals, there are significant differences between signals and traditional tasks such as image or text generation. Traditional GenAI models fail to generate outputs that align with RF signal characteristics, leading to hallucinations in wireless sensing.
%缩短前
% GenAI models can support wireless data augmentation, signal detection, and time-series prediction. However, applying GenAI models to generate radio frequency (RF) signals may lead to wireless hallucinations, such as the omission or misrepresentation of signal frequency characteristics or phase information. Additionally, generated signals may appear excessively smoothed or overly noisy in the spectrum. These issues stem from the difficulty that GenAI models face in capturing the dynamic temporal and complex-valued characteristics of RF signals.
GenAI supports wireless data augmentation, signal detection, and time-series prediction. However, generating radio frequency (RF) signals with GenAI can introduce wireless hallucinations, such as missing or misrepresenting signal frequency characteristics and phase information. Additionally, generated signals may appear excessively smoothed or overly noisy in the spectrum. These issues arise from GenAI's challenges in capturing the dynamic temporal and complex-valued nature of RF signals.
% 缩短前
% To address this, the authors in~\cite{chi2024rf} proposed a novel GenAI model based on time-frequency diffusion theory and a hierarchical diffusion transformer (HDT) to generate high-quality and complex time-series RF data, where the signal-to-noise ratio (SNR) is used to measure the hallucination effect in channel estimation tasks. First, noise and blurring operations are simultaneously introduced in both the time and frequency domains of the RF signals to achieve data augmentation. This ensures that the generated RF signals maintain high fidelity in both domains, with more accurate spectral details. Beyond data augmentation, the hierarchical transformer architecture introduced in the GDM decouples different types of noise (e.g., Gaussian noise and spectral blurring) introduced during the diffusion process. Each stage is specifically optimized for a particular type of noise, thereby improving generation accuracy. Compared to VAEs based method, this approach significantly enhances the detail of the generated signals, reducing wireless hallucination effect by $77.5\%$ in 5G frequency domain duplex (FDD) channel estimation.
To address this issue, the authors in~\cite{chi2024rf} proposed a GenAI model based on time-frequency diffusion theory and a hierarchical diffusion transformer (HDT) to generate high-quality RF time-series data. In channel estimation tasks, signal-to-noise ratio (SNR) is used to quantify hallucination effects. The model applies noise and blurring operations in both time and frequency domains for data augmentation, preserving spectral details and improving fidelity. Additionally, the hierarchical transformer in the GDM decouples different noise types (e.g., Gaussian noise, spectral blurring), optimizing each stage for a specific noise type to enhance accuracy. Compared to VAE-based methods, this approach improves signal detail and reduces wireless hallucinations by $77.5\%$ in 5G frequency division duplex (FDD) channel estimation.

\subsubsection{GenAI-enabled Radio Map Construction}

% By leveraging sparse sampling points to learn the spatial propagation characteristics of signals, GenAI models can generate high-resolution radio maps from sparse inputs. However, current GenAI-based generation methods often rely on single datasets, resulting in poor generalization capabilities and difficulty in adapting to multi-scenario, multi-frequency, and three-dimensional modeling requirements. Furthermore, if environmental information (e.g., terrain height or building distribution) is incomplete, GenAI models may produce inaccurate results, leading to hallucination phenomena.
% 缩短前
% GenAI models can generate high-resolution radio maps from sparse inputs. However, it can also cause wireless hallucinations, such as the inability of the generated radio maps to accurately reflect actual multipath effects and signal attenuation. It can also lead to the failure of two-dimensional radio maps to capture the signal propagation characteristics influenced by building heights, and received signal strength (RSS) values that violate physical constraints. These issues arise from limited dataset diversity and poor generalization ability of the models.
GenAI models can generate high-resolution radio maps from sparse inputs but may also introduce wireless hallucinations. For example, generated maps may fail to accurately capture multipath effects and signal attenuation. Additionally, two-dimensional radio maps may overlook signal propagation variations caused by building heights, and received signal strength might violate physical constraints. These issues stem from limited dataset diversity and the models’ poor generalization ability.
% 缩短前
% To address this, the authors in~\cite{10757328} released the \textit{SpectrumNet} dataset to effectively reduce the wireless hallucination effect of the GenAI models in radio map construction from the data augmentation perspective. By considering propagation effects such as signal transmission, refraction, and diffraction, ray tracing was utilized to simulate wireless signal propagation paths. Additionally, real-world geographic information and climate models provided by the ITU were used to construct a comprehensive radio map dataset. By testing the generative models across multiple dimensions, including different terrains, altitudes, and frequencies, it is demonstrated that the proposed \textit{SpectrumNet} dataset is essential for training radio map generation models with the ability to resist wireless hallucinations.
% Furthermore, the authors of~\cite{sun2024generative} employed IAI and LLM methods to guide the diffusion model in dynamically adjusting design parameters, effectively reducing hallucination effects in UAV-assisted spectrum map construction.
To address this issue, the authors in~\cite{10757328} released the \textit{SpectrumNet} dataset to effectively reduce the wireless hallucination effect of the GenAI models in radio map construction from the data augmentation perspective. Ray tracing was employed to simulate wireless signal propagation paths by considering transmission, refraction, and diffraction effects. Additionally, real-world geographic data and climate models from the ITU were incorporated to enhance dataset comprehensiveness. Through multi-dimensional testing across various terrains, altitudes, and frequencies, it is demonstrated that the proposed \textit{SpectrumNet} dataset is essential for training radio map generation models with the ability to resist wireless hallucinations.

\subsection{Data Link Layer Perspective}

\subsubsection{GenAI-enabled Semantic Communication}

GenAI can generate background knowledge tailored to specific communication scenarios, enhancing semantic communication adaptability. However, it may introduce wireless hallucinations due to information redundancy and weak contextual relevance. These issues include the semantic encoder embedding irrelevant or ambiguous information during encoding and the receiver’s decoder failing to accurately reconstruct the original semantic content, leading to deviations from its original intent. To address this, the authors in~\cite{10433140} proposed the Gen-SC framework, incorporating adversarial training as a model-level solution. This mechanism alternates optimization between a generator and a discriminator to improve GenAI performance. A context regulator dynamically adjusts input background data, ensuring generated samples align with semantic communication needs and reducing hallucinations. Hallucination is quantified using sentence similarity metrics, with higher scores indicating better semantic consistency. Additionally, a neural network-based discriminator evaluates topic and semantic similarity between generated and real samples. Simulations show that Gen-SC reduces hallucinations by $9.76\%$ compared to models without adversarial training.

\subsubsection{GenAI-enabled Non-orthogonal Multiple Access}

With efficient codebook design and optimized encoding/decoding, GenAI enhances NOMA network performance. However, their application in NOMA, including sparse code multiple access (SCMA), may introduce hallucinations, such as generating codebooks that fail to meet network requirements and excessive decoding latency, compromising reliability. For example, a GenAI model trained on outdoor line-of-sight (LoS) data may generate beam codebooks unsuitable for indoor non-line-of-sight (NLoS) environments. These issues stem from training challenges including vanishing gradients, non-convergence, and mode collapse, especially in GAN-based models. To address this, the authors in\cite{9925625} integrated Wasserstein distance (WD)-based GANs with autoencoders (AE) to improve decoding in 6G SCMA systems. The WD-based loss function stabilizes gradient feedback, mitigating mode collapse, while embedding the AE’s encoder within GANs ensures generated codebooks are both theoretically well-distributed and practically decodable, enhancing robustness.

\subsection{Network Layer Perspective}

\subsubsection{GenAI-enabled Wireless QoE Improvement}

GenAI plays a key role in wireless network resource management by capturing high-dimensional data in complex networks. However, challenges in obtaining subjective QoE feedback and limited model adaptability can cause wireless hallucinations in AIGC service distribution, such as generated images misaligned with user intent or aesthetics, logical inconsistencies, and poor adaptation to dynamic conditions. To address this, the authors in~\cite{du2023user} proposed a dynamic feedback and IAI-based QoE-aware optimization framework. RL enhances GDM’s adaptability to changing environments, improving optimization efficiency. Additionally, an LLM-based IAI method simulates users' Big Five personality traits, using prompt engineering to generate subjective ratings that guide the RL-assisted GDM in producing QoE-optimized images. This framework improved QoE by $10\%\sim15\%$, significantly reducing wireless hallucinations in GenAI-driven optimization.

\subsubsection{GenAI-enabled Security Communication}

GenAI enhances wireless security by autonomously learning threat distributions and countering attacks. However, it can introduce wireless hallucinations, such as misclassifying legitimate signals as attacks, leading to unnecessary capacity reduction. These issues stem from limited adaptability of models and imbalanced or incomplete training data. To address this, the authors in~\cite{10839236} proposed an MoE-driven GenAI model for security optimization. The MoE framework decomposes multi-objective optimization into sub-tasks, each handled by expert models focusing on metrics including secrecy rate and secure energy efficiency. Additionally, a dynamic gating mechanism selects the optimal expert based on input states, improving resource allocation and computational efficiency.

\section{Case Study: Wireless Hallucination Mitigation Method for GDM-Based Channel Estimation}

\subsection{Design Overview}

\begin{figure*}
    \centering
    \includegraphics[width=0.8\linewidth]{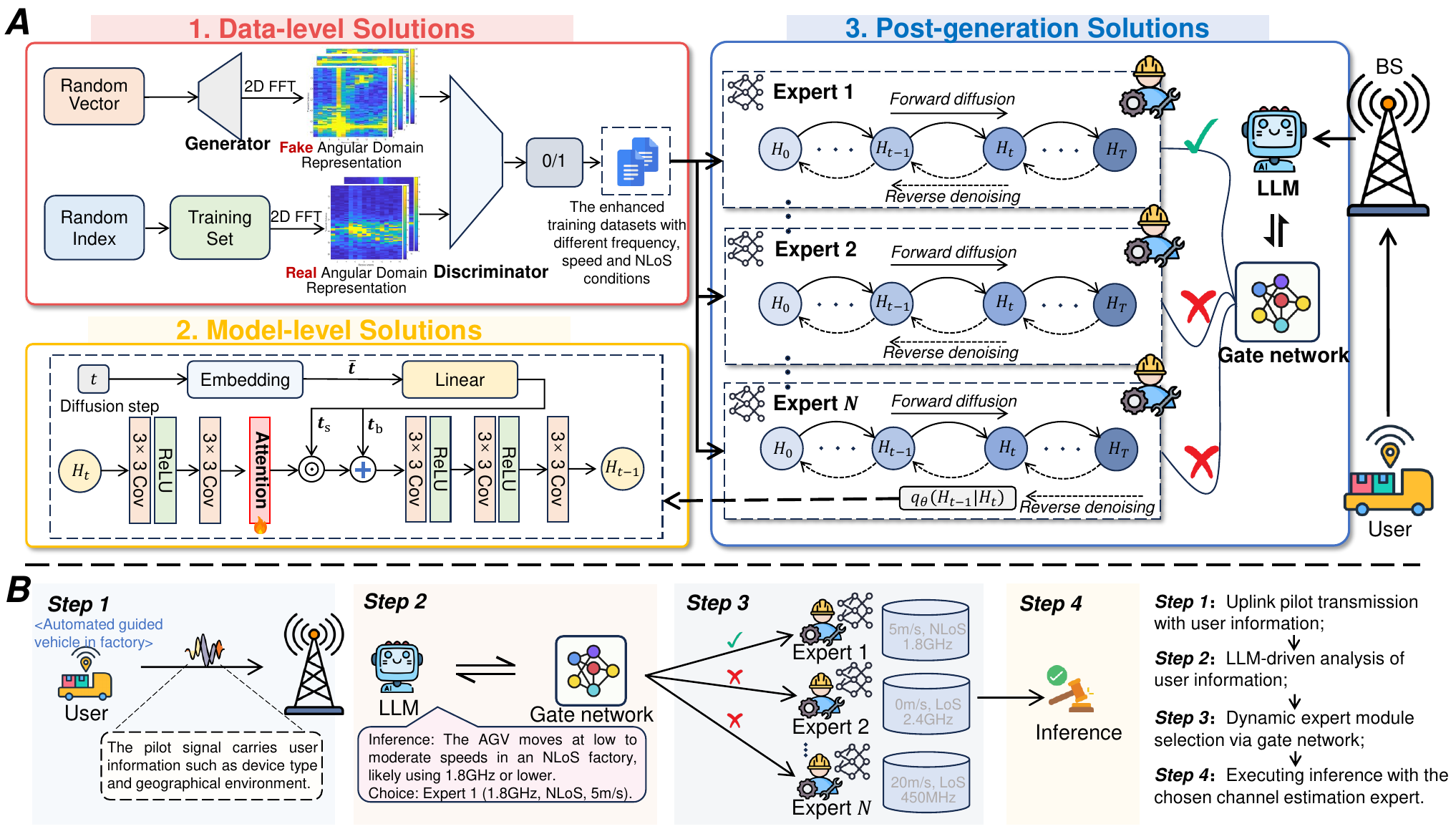}
    % 缩短前
    % \caption{\small{Part A illustrates the proposed integrated mitigation strategy to reduce wireless hallucination in channel estimation. In data-level solutions, a GAN is employed to enhance uneven training data. In model-level solutions, the attention mechanism is introduced to capture latent features of channel distribution effectively. In post-generation solutions, an LLM-enhanced MoE architecture is utilized to select models aligned with user requirements for channel estimation. Part B describes the workflow of the proposed channel estimation framework: Upon receiving pilot signals carrying user information, the LLM analyzes and infers user requirements, subsequently selecting the most appropriate expert to execute channel estimation.}}
    \caption{\small{Part A illustrates the proposed integrated solution for reducing wireless hallucination in channel estimation. At the data level, a GAN enhances uneven training data. At the model level, an attention mechanism captures latent channel distribution features. At the post-generation level, an LLM-enhanced MoE selects models that align with user requirements for for channel estimation. Part B outlines the workflow: Upon receiving pilot signals, the LLM analyzes and infers user requirements, subsequently selecting the suitable expert for channel estimation.}}
    \label{figure_case_study}
\end{figure*}

We present a case study on uplink channel estimation using GDMs and propose an integrated solution to mitigate hallucinations in estimation models. This study provides a practical example of integrating data-level, model-level, and post-generation solutions into a unified framework, an approach that has not been comprehensively explored in previous research. In data-level solutions, we preprocess the raw channel dataset to classify different data types, and then we use a GAN model for data augmentation. In model-level solutions, we develop the attention mechanism by enhancing the GDM-based estimation structure to reduce misleading feature influence. In post-generation solutions, a mobile edge MoE architecture is designed, integrating GDM experts for channel estimation under various conditions. An LLM-enhanced gate network dynamically selects the most suitable expert to reduce hallucinations in channel estimation. The proposed solution and workflow of channel estimation are shown in Fig.~\ref{figure_case_study}.

\subsection{Integrated Mitigation Strategy for Wireless Hallucination}

% 缩短前
% \subsubsection{Data-level Solution}First, we construct a channel dataset \textit{A} using QuaDriGa~\cite{quadriga2025}, incorporating various combinations of carrier frequency, mobility speed, and LoS/NLoS conditions, with a total of 10,000 samples. Notably, channel data under NLoS conditions are more challenging to obtain, resulting in fewer NLoS samples compared to LoS samples. To ensure a fair comparison, we also construct a mixed channel dataset \textit{B}, which does not distinguish between data types and contains the same number of 10,000 samples as a baseline. To address the data imbalance, a generative adversarial network (GAN) is employed, where the generator and discriminator are trained using an adversarial training mechanism to generate realistic synthetic channel data for the underrepresented NLoS condition. This data augmentation approach enhances the diversity of the dataset, thereby improving the generalization capability of the channel estimation model across different scenarios.
\subsubsection{Data-level Solution}First, we construct a channel dataset \textit{A} using QuaDriGa~\cite{quadriga2025}, incorporating various carrier frequencies, mobility speeds, and LoS/NLoS conditions, totaling 10,000 samples. Notably, NLoS channel data are harder to obtain, leading to fewer NLoS samples than LoS samples. To enable a fair comparison, we also create a mixed channel dataset \textit{B}, which does not differentiate data types but maintains the same 10,000-sample size as a baseline. To address data imbalance, we employ GANs, where the generator and discriminator undergo adversarial training to synthesize realistic NLoS channel data. This data augmentation strategy enhances dataset diversity, improving the generalization ability of the channel estimation model across different scenarios. The released datasets can be found in \url{https://github.com/wang104225/WirelessHallu}.

% 缩短前
% \subsubsection{Model-level Solution}To further mitigate the hallucination effect caused by the randomness in the GenAI generation process, we design an attention-enhanced diffusion model to improve channel estimation accuracy. Specifically, the attention layer is introduced between the first convolutional block and subsequent layers to capture global dependencies across the input channel features. The feature map is projected into query, key, and value matrices, where attention scores are calculated to assign weights to the value matrix, enriching the representation with global context before passing it to the $3\times3$ convolutional layers. The integration of the attention layer enhances the robustness of GenAI by capturing interdependencies among spatial and angular features in challenging wireless environments.
\subsubsection{Model-level Solution}To further reduce hallucinations caused by the randomness in GenAI-generated outputs, we design an attention-enhanced diffusion model to improve channel estimation accuracy. Specifically, the attention layer is introduced between the first convolutional block and subsequent layers to capture global dependencies across input channel features. The feature map is projected into query, key, and value matrices, where attention scores are calculated to assign weights to the value matrix, enriching the representation with global context before passing it to the $3\times3$ convolutional layers. By integrating the attention mechanism, the model enhances robustness by capturing spatial and angular interdependencies, particularly in challenging wireless environments.

\subsubsection{Post-generation Solution}We introduce an LLM-enhanced MoE approach to determine the applicability of expert models for channel estimation across different users. In the architecture, diffusion models at the base station (BS) serve as expert modules, while the LLM-enhanced gating network determines which expert should be activated for channel estimation. Leveraging its powerful analytical capabilities, the LLM deployed at the mobile edge can infer the necessary expert module directly from user state information without requiring additional training. Moreover, the user state information, including environmental conditions, radio frequency, and mobility speed, collectively influences the activation decisions made by the LLM.

\subsection{Experimental Results}

% 缩短前
% \subsubsection{Experimental Configuration}Our experiment is implemented with torch 2.5.1 and Python 3.12.8. The proposed model is based on the GDM framework, which is trained on a Ubuntu server with a 2.10 GHz Intel Xeon Gold 5218R CPU, 40 cores, and 503 GB memory, along with an NVIDIA A100 80 GB GPU. The number of diffusion steps is set to 100. We deploy Llama 3 LLMs with 7 billion parameters to support the gate network. In addition, we adopt the following baselines: \textbf{Hallucination Strategy}, which is the diffusion model-based channel estimation method without any hallucination mitigation strategy; \textbf{Hybrid Strategy-w/o Attention}, which employs GAN-based data augmentation during the training phase but does not incorporate the attention mechanism; \textbf{Hybrid Strategy-w/o LLM}, which uses the random expert selection strategy during the inference phase.
\subsubsection{Experimental Configuration}Our experiment is implemented using Torch 2.5.1 and Python 3.12.8. The proposed model is built on the GDM framework and trained on an Ubuntu server equipped with a 2.10 GHz Intel Xeon Gold 5218R CPU (40 cores, 503 GB RAM) and an NVIDIA A100 80 GB GPU. The number of diffusion steps is set to 60. And Llama 3 LLMs with 7 billion parameters are deployed to support the gate network.
\subsubsection{Experimental Results}
\begin{figure*}[t]
\centering
    \begin{subfigure}[b]{0.35\textwidth}
            \includegraphics[width=\textwidth]{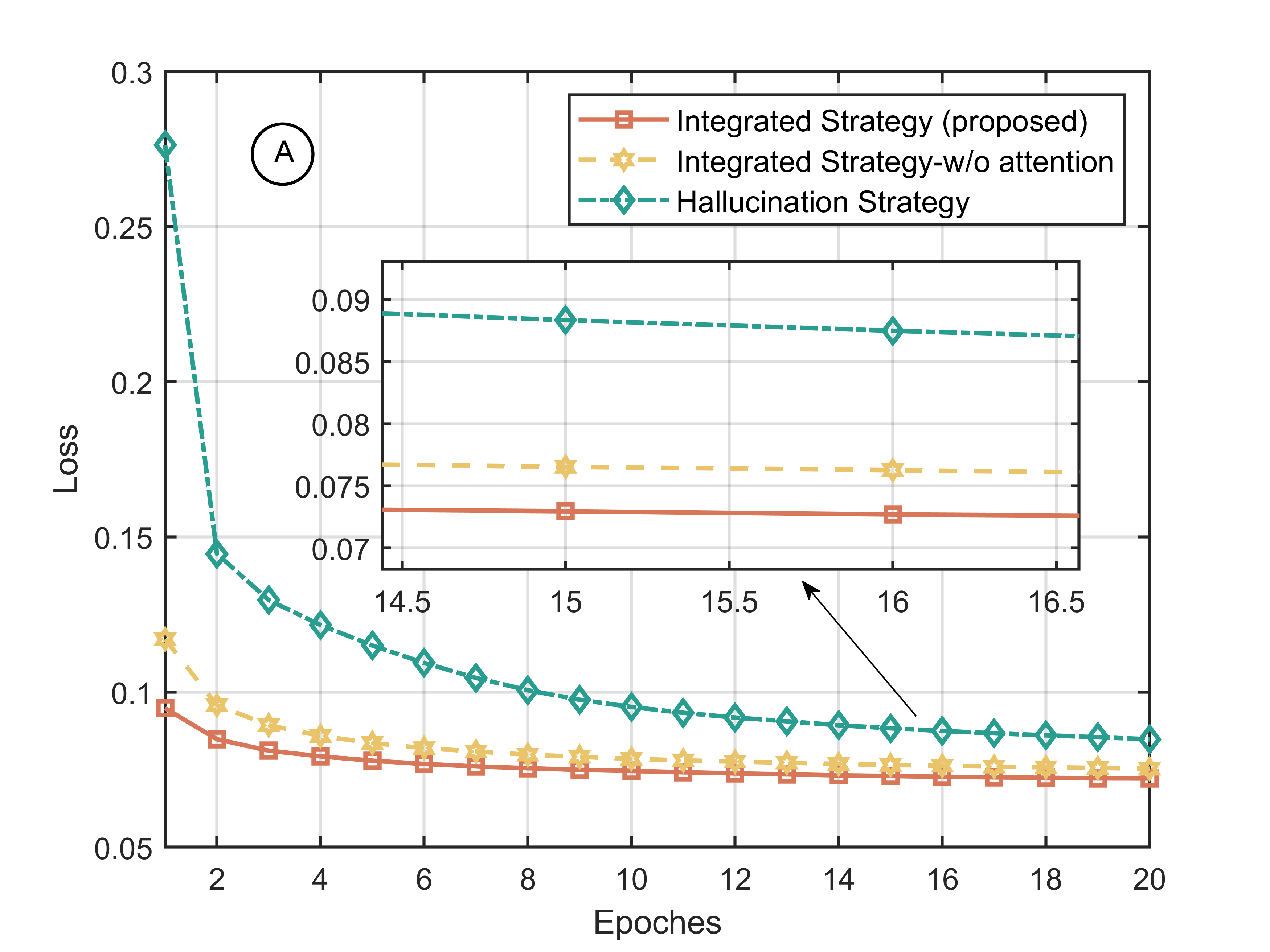}
            %\caption{}
            \label{loss_convergence}
    \end{subfigure}
    \hfil
    \begin{subfigure}[b]{0.35\textwidth}
            \includegraphics[width=\textwidth]{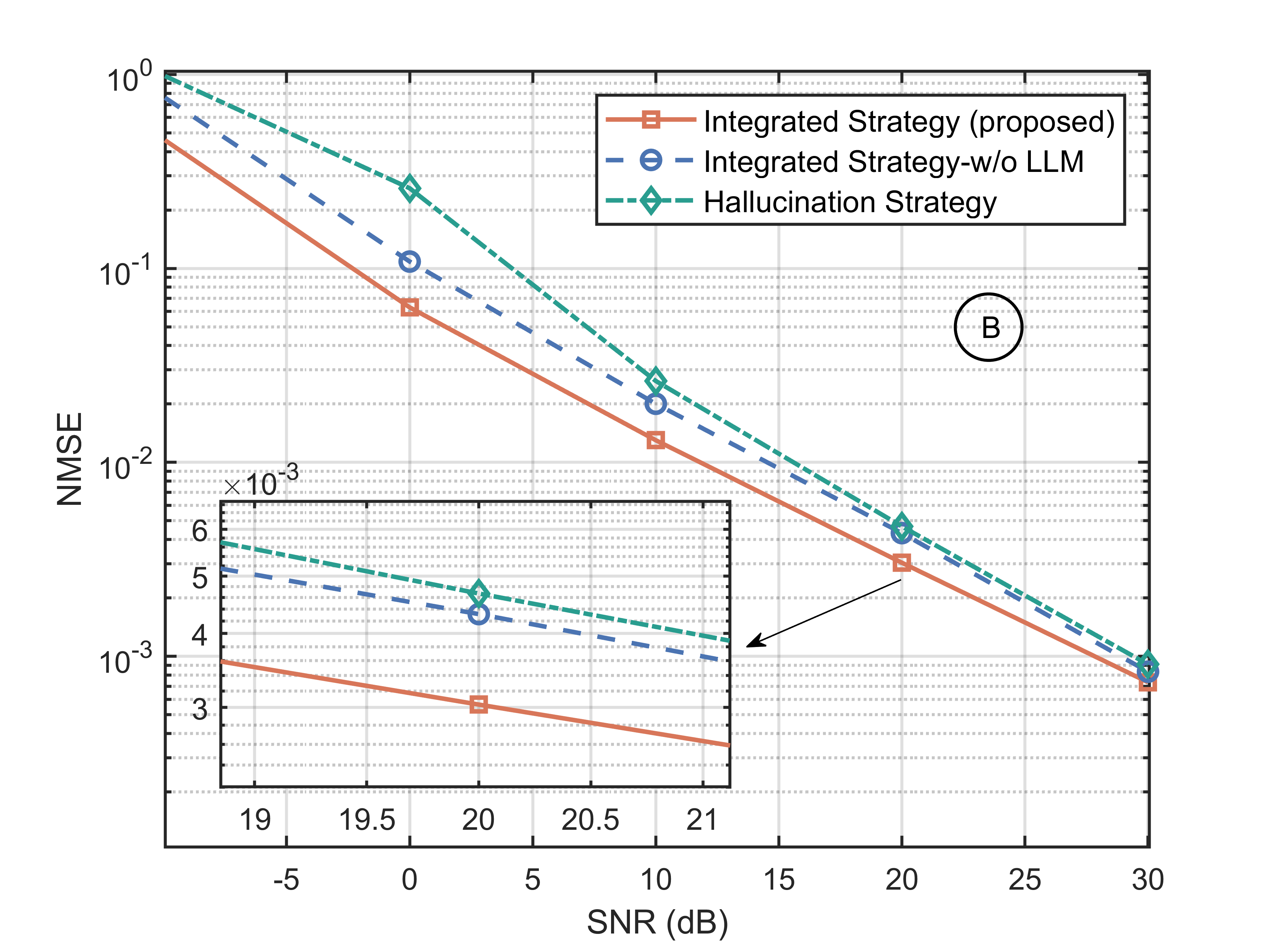}
            %\caption{NMSE performance of the proposed integrated mitigation strategy and other baselines over SNR.}
            \label{NMSE_SNR}
    \end{subfigure}
\caption{\small{Experimental results. (\textbf{A}): The training loss curves of the proposed integrated mitigation strategy and other baselines over epochs.} (\textbf{B}): NMSE performance of the proposed integrated mitigation strategy and other baselines over SNR. Note that we adopt the following baselines: \textbf{Hallucination Strategy}, a diffusion-based channel estimation method without any hallucination mitigation; \textbf{Integrated Strategy-w/o Attention}, which employs GAN-based data augmentation training but excludes the attention mechanism; \textbf{Integrated Strategy-w/o LLM}, which uses random expert selection during inference.}
\label{experiment_results}
\end{figure*}

Fig.~\ref{experiment_results}(a) presents the training loss of the proposed integrated strategy for hallucination alongside baselines. Due to data augmentation and the attention mechanism, the proposed strategy converges to a lower loss value than the hallucination strategy, demonstrating its effectiveness in reducing hallucinations. Additionally, it outperforms the integrated strategy without attention, confirming that the attention mechanism in GDMs enhances the model’s ability to capture fine-grained details. We further evaluate the NMSE of the proposed strategy across different SNR levels, as shown in Fig.~\ref{experiment_results}(b). The integrated strategy achieves lower NMSE across all SNR levels, with a 0.19 reduction at 0 dB SNR compared to the hallucination strategy. Notably, the proposed method demonstrates a greater hallucination reduction effect in low-SNR regions than in high-SNR regions. This is because noise has a stronger impact on model generation in low-SNR scenarios, making noise-guided generation more prone to unrealistic channel estimates. Furthermore, the proposed strategy outperforms the version without LLMs, highlighting that LLM integration effectively selects the most suitable expert for channel estimation, further reducing wireless hallucinations.

\section{CONCLUSION}

% 缩短前
% This paper examined the potential wireless hallucinations in GenAI-enabled communications. First, we introduced the concept of wireless hallucinations and systematically analyze the GenAI workflow to summarize their causes. Then, we investigated hallucination mitigation solutions at the data, model, and post-generation levels. Furthermore, we provided a comprehensive review of existing wireless hallucinations in GenAI-enabled communications and their corresponding mitigation strategies. Finally, we proposed an effective integrated strategy to reduce wireless hallucinations in GDM-based channel estimation. Specifically, we published a channel estimation hallucination dataset and employed GANs to enhance training data. Additionally, attention mechanisms and LLMs are integrated into both the training and inference phases, reducing NMSE represented wireless hallucinations by $0.19$.
In this paper, we investigated wireless hallucinations in GenAI-enabled communications. We first defined wireless hallucinations and analyzed the GenAI workflow to identify their causes. Then, we explored mitigation strategies at the data, model, and post-generation levels. Additionally, we reviewed existing cases of wireless hallucinations and corresponding countermeasures. Finally, we proposed an integrated strategy for reducing hallucinations in GDM-based channel estimation. Specifically, we published a hallucination dataset and use GANs to enhance training data. Attention mechanisms and LLMs are incorporated into both training and inference, reducing NMSE-based hallucinations by $0.19$.

%\printbibliography % 对应 package:biblatex

% \bibliographystyle{ieeetr} % 会导致number 和 month 只显示 month
\bibliographystyle{IEEEtran}
\bibliography{manuscript.bbl}

\begin{thebibliography}{10}
\providecommand{\url}[1]{#1}
\csname url@samestyle\endcsname
\providecommand{\newblock}{\relax}
\providecommand{\bibinfo}[2]{#2}
\providecommand{\BIBentrySTDinterwordspacing}{\spaceskip=0pt\relax}
\providecommand{\BIBentryALTinterwordstretchfactor}{4}
\providecommand{\BIBentryALTinterwordspacing}{\spaceskip=\fontdimen2\font plus
\BIBentryALTinterwordstretchfactor\fontdimen3\font minus
  \fontdimen4\font\relax}
\providecommand{\BIBforeignlanguage}[2]{{%
\expandafter\ifx\csname l@#1\endcsname\relax
\typeout{** WARNING: IEEEtran.bst: No hyphenation pattern has been}%
\typeout{** loaded for the language `#1'. Using the pattern for}%
\typeout{** the default language instead.}%
\else
\language=\csname l@#1\endcsname
\fi
#2}}
\providecommand{\BIBdecl}{\relax}
\BIBdecl

\bibitem{khoramnejad2025generative}
F.~Khoramnejad and E.~Hossain, ``{Generative AI for the Optimization of
  Next-generation Wireless Networks: Basics, State-of-the-art, and Open
  Challenges},'' \emph{IEEE Communications Surveys \& Tutorials}, 2025.

\bibitem{chakraborty2025hallucination}
N.~Chakraborty, M.~Ornik, and K.~Driggs-Campbell, ``{Hallucination Detection in
  Foundation Models for Decision-making: A Flexible Definition and Review of
  The State of The Art},'' \emph{ACM Computing Surveys}, 2025.

\bibitem{huang2024survey}
L.~Huang, W.~Yu, W.~Ma, W.~Zhong, Z.~Feng, H.~Wang, Q.~Chen, W.~Peng, X.~Feng,
  B.~Qin \emph{et~al.}, ``{A Survey on Hallucination in Large Language Models:
  Principles, Taxonomy, Challenges, and Open Questions},'' \emph{ACM
  Transactions on Information Systems}, 2024.

\bibitem{chi2024rf}
G.~Chi, Z.~Yang, C.~Wu, J.~Xu, Y.~Gao, Y.~Liu, and T.~X. Han, ``{RF-Diffusion:
  Radio Signal Generation via Time-Frequency Diffusion},'' in \emph{Proceedings
  of the 30th Annual International Conference on Mobile Computing and
  Networking}, 2024, pp. 77--92.

\bibitem{10874185}
M.-H. Wu, H.-Y. Chen, T.-W. Yang, C.-C. Hsu, C.-W. Huang, and C.-F. Chou,
  ``{Intelligent Reflecting Surface-Assisted Millimeter Wave Communications:
  Cross Attention-aided Variational Autoencoder-based Precoding Design},''
  \emph{IEEE Transactions on Cognitive Communications and Networking}, pp.
  1--1, 2025.

\bibitem{9925625}
L.~Miuccio, D.~Panno, and S.~Riolo, ``{A Flexible Encoding/Decoding Procedure
  for 6G SCMA Wireless Networks via Adversarial Machine Learning Techniques},''
  \emph{IEEE Transactions on Vehicular Technology}, vol.~72, no.~3, pp.
  3288--3303, 2023.

\bibitem{wang2024generative}
X.~Wang, H.~Du, D.~Niyato, L.~Zhou, L.~Feng, Z.~Yang, F.~Zhou, and W.~Li,
  ``{Generative AI Enabled Matching for 6G Multiple Access},'' \emph{arXiv
  preprint arXiv:2411.04137}, 2024.

\bibitem{ji2023survey}
Z.~Ji, N.~Lee, R.~Frieske, T.~Yu, D.~Su, Y.~Xu, E.~Ishii, Y.~J. Bang,
  A.~Madotto, and P.~Fung, ``{Survey of Hallucination in Natural Language
  Generation},'' \emph{ACM Computing Surveys}, vol.~55, no.~12, pp. 1--38,
  2023.

\bibitem{10757328}
S.~Zhang, S.~Jiang, W.~Lin, Z.~Fang, K.~Liu, H.~Zhang, and K.~Chen,
  ``{Generative AI on SpectrumNet: An Open Benchmark of Multiband 3D Radio
  Maps},'' \emph{IEEE Transactions on Cognitive Communications and Networking},
  pp. 1--1, 2024.

\bibitem{10433140}
F.~Zhao, Y.~Sun, L.~Feng, L.~Zhang, and D.~Zhao, ``{Enhancing Reasoning Ability
  in Semantic Communication Through Generative AI-Assisted Knowledge
  Construction},'' \emph{IEEE Communications Letters}, vol.~28, no.~4, pp.
  832--836, 2024.

\bibitem{zhang2024emergence}
H.~Zhang, J.~Zhou, Y.~Lu, M.~Guo, P.~Wang, L.~Shen, and Q.~Qu, ``{The Emergence
  of Reproducibility and Consistency in Diffusion Models},'' in
  \emph{Forty-first International Conference on Machine Learning}, 2024.

\bibitem{10839236}
C.~Zhao, H.~Du, D.~Niyato, J.~Kang, Z.~Xiong, D.~I. Kim, X.~S. Shen, and K.~B.
  Letaief, ``{Enhancing Physical Layer Communication Security through
  Generative AI with Mixture of Experts},'' \emph{IEEE Wireless
  Communications}, pp. 1--9, 2025.

\bibitem{du2023user}
H.~Du, R.~Zhang, D.~Niyato, J.~Kang, Z.~Xiong, S.~Cui, X.~Shen, and D.~I. Kim,
  ``{User-centric interactive AI for distributed diffusion model-based
  AI-generated content},'' \emph{arXiv preprint arXiv:2311.11094}, 2023.

\bibitem{quadriga2025}
\BIBentryALTinterwordspacing
F.~H.~H. Institute. (2025) {QuaDRiGa - Quasi Deterministic Radio Channel
  Model}. [Online]. Available: \url{https://quadriga-channel-model.de/}
\BIBentrySTDinterwordspacing

\end{thebibliography}

% \section*{BIOGRAPHIES}  %不带序号
% \textbf{Xudong Wang} is a PhD candidate at Beijing University of Posts and Telecommunications, China. \textbf{Jiacheng Wang} is a research fellow at Nanyang Technological University, Singapore. \textbf{Lei Feng} is an associate professor with Beijing University of Posts and Telecommunications, China. \textbf{Dusit Niyato} is a chair professor with Nanyang Technological University, Singapore. \textbf{Ruichen Zhang} is a research fellow at Nanyang Technological University, Singapore. \textbf{Jiawen Kang} is a professor with Guangdong University of Technology, China. \textbf{Zehui Xiong} is an assistant professor with Singapore University of Technology and Design, Singapore. \textbf{Hongyang Du} is an assistant professor with The University of Hong Kong, Hong Kong. \textbf{Shiwen Mao} is a professor with Auburn University, Auburn, USA.

%\vfill

\end{document}